\documentclass{elsart}

\usepackage{graphicx}
\usepackage{subfigure}
\usepackage{amssymb}
\usepackage[dvips]{color}

\begin{document}

\begin{frontmatter}

\title{Charge order, orbital order, and electron localization in the 
       Magn\'{e}li phase $ {\rm \bf Ti_4O_7} $}

\author{V.\ Eyert\corauthref{corauth}}, 
\corauth[corauth]{Corresponding author. fax: +49 821 598 3262} 
\ead{eyert@physik.uni-augsburg.de}
\author{U.\ Schwingenschl\"ogl}, 
\author{U.\ Eckern}, 

\address{Institut f\"ur Physik, Universit\"at Augsburg, 
         86135 Augsburg, Germany}

\begin{abstract}
The metal-insulator transition of the Magn\'{e}li phase $ {\rm Ti_4O_7} $ 
is studied by means of augmented spherical wave (ASW) electronic structure 
calculations as based on density functional theory and the local density 
approximation. The results show that the metal-insulator transition arises 
from a complex interplay of charge order, orbital order, and singlet 
formation of those Ti $ 3d $ states which mediate metal-metal bonding 
inside the four-atom chains characteristic of the material. $ {\rm Ti_4O_7} $ 
thus combines important aspects of $ {\rm Fe_3O_4} $ and $ {\rm VO_2} $. 
While the charge ordering closely resembles that observed at the 
Verwey transition, the orbital order and singlet formation 
appear to be identical to the mechanisms driving the metal-insulator 
transition of vanadium dioxide. 
\end{abstract}

\end{frontmatter}

\section{Introduction}

Since the seminal work of Morin \cite{morin59} the metal-insulator 
transitions of the early transition-metal oxides are motivating 
ongoing research. In particular, showing first-order transitions with 
a conductivity change of several orders of magnitude, $ {\rm VO_2} $ 
and $ {\rm V_2O_3} $ are regarded as prototypical materials of this 
class. Recent theoretical work has revealed a strong influence via 
electron-phonon coupling of the structural degrees of freedom on the 
electronic properties of $ {\rm VO_2} $ and neighbouring rutile-type 
dioxides \cite{wentz94a,eyert02b,eyert00,eyert02a}. In contrast, the 
corundum-type sesquioxide has turned out to be subject to strong 
electronic correlations \cite{ezhov99,held01}, as seems to be the 
case also for $ {\rm Ti_2O_3} $ \cite{poteryaev03}. The relative 
importance of these mechanisms as well as their dependence on crystal 
structure and band filling are still unclear. 

In this situation the Magn\'{e}li phases offer the opportunity of a 
more comprehensive understanding \cite{andersson57}. These phases 
form a homologous series of the kind $ {\rm M_nO_{2n-1}} $ 
($ {\rm M = Ti, V} $) and are particularly suited for studying the 
differences in crystal structures and electronic properties between 
the end members $ {\rm MO_2} $ ($ {\rm n \to \infty} $) and 
$ {\rm M_2O_3} $ ($ {\rm n = 2} $). This is due to the variation in 
$ d $ band occupation across the series as well as the fact that, 
as the general formula $ {\rm M_nO_{2n-1} = M_2O_3 + (n-2) MO_2} $ 
suggests, the crystal structures can be viewed as rutile-type 
slabs of infinite extension and different thickness, separated by 
shear planes with a corundum-like atomic arrangement \cite{andersson57}. 
While in the rutile-type regions the characteristic metal-oxygen 
octahedra are coupled via edges, the shear planes have face-sharing 
octahedra as in the corundum structure. We have recently started a 
systematic investigation of the vanadium Magn\'{e}li phases 
\cite{schwingen03a,schwingen03b,dissudo,schwingen04}; here we report 
on the first results for the titanium series. 

The Magn\'{e}li phase $ {\rm Ti_4O_7} $ experiences a metal-insulator 
transition at 154 K, which is followed by an insulator-insulator 
transition at 130 K with a thermal hysteresis of about 12 K 
\cite{bartholomew69,schlenker74,lakkis76}. Both transitions are first-order 
\cite{schlenker74,lakkis76,marezio72} and connected with a conductivity 
change of three orders of magnitude each \cite{bartholomew69}. While a 
sharp decrease of the magnetic susceptibility has been reported for the 
154 K transition, which has been interpreted as a change from Pauli 
paramagnetism to van Vleck-like behaviour \cite{lakkis76}, there is no 
considerable change at lower temperatures \cite{bartholomew69,mulay70}. 
In contrast, whereas the transition at 154 K goes along with increase in 
volume \cite{marezio72}, a distortion of the crystal structure had been 
first observed only for the second transition \cite{bartholomew69}. From 
single-crystal x-ray diffraction data taken at 298, 140, and 120 K, 
Marezio {\em et al.}\ reported  differentiation of the Ti-O distances 
on going to the low-temperature phase, suggesting the formation of 
$ {\rm Ti^{3+}} $ and $ {\rm Ti^{4+}} $ sites, which each occupy half of 
the characteristic four-atom chains and which contrast the average 
$ {\rm Ti^{3.5+}} $ valence observed at high temperatures  
\cite{marezio72,marezio71,marezio73}. 
In addition, the $ {\rm Ti^{3+}} $ sites form short metal-metal bonds 
indicating singlet formation and explaining the changes in resistivity 
and magnetic susceptibility. At the same time the unpaired $ {\rm Ti^{4+}} $ 
and, to a lesser degree, the $ {\rm Ti^{3+}} $ atoms are displaced 
perpendicular to the chain axis off the center of the surrounding 
octahedron towards one of the oxygen atoms 
\cite{marezio72,marezio71,marezio73}. This pattern is similar to the 
situation found for Cr-doped $ {\rm VO_2} $, where half of the chains 
pair and atoms on the remaining chains display a zigzag-like off-center 
displacement \cite{eyert02b}. Finally, Marezio {\em et al.}\ reported 
large thermal displacements for the intermediate phase, suggesting that 
the structural changes are associated with the metal-insulator 
transition at 154 K but still lack long-range order 
\cite{marezio72,marezio73}. The latter sets in at the lower transition, 
which is thus regarded as a disorder-order transition 
\cite{marezio72,marezio73} as has been confirmed by specific-heat and 
EPR data \cite{schlenker74,lakkis76}. These results led to interprete 
the intermediate phase as a bipolaron liquid 
\cite{schlenker74,lakkis76,schlenker80}. Yet, refined determination of 
the crystal structure revealed a fivefold superstructure for the 
intermediate phase, which shows the same structural characteristics as 
the low-temperature phase, and thus questioned understanding the 
conductivity in terms of independent mobile bipolarons \cite{lepage84}. 
Recent photoemission and x-ray absorption measurements showed only 
minor effects of the insulator-insulator transition on the electronic 
properties but revealed drastic changes at 154 K 
\cite{abbate95,kobayashi02}. The latter include significant down- and 
upshift of the occupied and unoccupied Ti $ 3d $ $ t_{2g} $ states near 
the Fermi energy. In contrast, O $ 2p $ states extending from $ -9 $ 
to $ -4 $\,eV remain essentially unaffected by the transition. 

Understanding of the physical properties of the Magn\'{e}li phases is 
greatly facilitated by an alternative unified representation of their 
crystal structures 
\cite{schwingen03a,schwingen03b,dissudo,schwingen04,marezio72,marezio73}. 
It is based on the notion of a hexagonal closed-packed array of oxygen atoms 
forming a regular 3D network of octahedra. Apart from a slightly different 
buckling, this network is the same for all Magn\'{e}li phases including 
the rutile-type dioxide and the corundum-type sesquioxide. Differences 
between the compounds $ {\rm M_nO_{2n-1}} $ arise from filling the 
octahedra with metal atoms. Filled octahedra form chains of length 
$ {\rm n} $ parallel to the pseudorutile $ c_{\rm prut} $ axis, 
followed by $ {\rm n-1} $ empty sites. While in the rutile structure 
these chains have infinite length, they comprise just two filled 
octahedra in the corundum structure, where they are perpendicular to 
the hexagonal $ c $ axis. Within the crystal, the metal chains are 
arranged in two different types of layers, which are interlaced by 
oxygen layers and alternate along $ a_{\rm prut} $. This gives rise to 
the aforementioned two types of chains, labelled (a) and (b). In 
$ {\rm Ti_4O_7} $, they comprise atoms Ti1/Ti3 and Ti2/Ti4, respectively 
(see Fig.\ 1 of Ref.\ \cite{schwingen03b}, where this situation has been 
sketched for the isostructural $ {\rm V_4O_7} $).
Due to alternation of the layers and relative shifts of the four-atom 
chains within the layers the end atoms Ti1 and Ti2 are found on top of 
each other (note that in the notation by Marezio and coworkers 
\cite{marezio72,marezio73} the chain center and end atoms are reversed). 
In the sesquioxide, these two atoms are usually designated 
as the $ c $-axis pair. While neighbouring octahedra share faces along 
$ a_{\rm prut} $ and $ b_{\rm prut} $, metal-metal bonding along all 
other directions within the layers is via edges \cite{eyert02b}. As a 
consequence, the atomic arrangement near the chain ends is corundum-like, 
whereas the chain centers correspond to the rutile-type regions. By 
virtue of the just sketched representation of the crystal structures it 
is possible to refer the symmetry components of the Ti $ 3d $ orbitals 
of all compounds to a common local coordinate system \cite{eyert02b}. 
In this system the $ z $- and $ x $-axes of the local coordinate system 
are parallel to the apical axis of the local octahedron and the pseudorutile 
$ c_{\rm prut} $ axis, respectively.

In this letter we report on electronic structure calculations for the 
Magn\'{e}li phase $ {\rm Ti_4O_7} $ using the crystal structures of 
both the room-temperature and the low-temperature phase. 
Our calculations reveal i) rather isotropic occupation of the Ti 
$ 3d $ $ t_{2g} $ states for the room-temperature structure, 
ii) significant electron transfer from one half of the chains to the other 
in the low-temperature structure, iii) orbital order at the $ d^1 $ 
chains coming with strong metal-metal dimerization, and iv) strong 
electron localization, which leads the way to the observed metal-insulator 
transition.

\section{Methodology}

The calculations were performed using the scalar-relativistic augmented  
spherical wave (ASW) method \cite{wkg,revasw}. In order to represent the 
correct shape of the crystal potential in the large voids of the open 
crystal structure, additional augmentation spheres were inserted. Optimal 
augmentation sphere positions as well as radii of all spheres were
automatically generated by the sphere geometry optimization (SGO) 
algorithm \cite{eyert98b}.
Self-consistency was achieved by an efficient algorithm for convergence
acceleration \cite{mixpap}. Brillouin zone sampling was done using an
increased number of $ {\bf k} $-points ranging from 108 to 2048 points 
within the irreducible wedge.

\section{Results and Discussion}

\begin{figure}[htb]
\centering 
\subfigure{\includegraphics[width=0.48\textwidth]{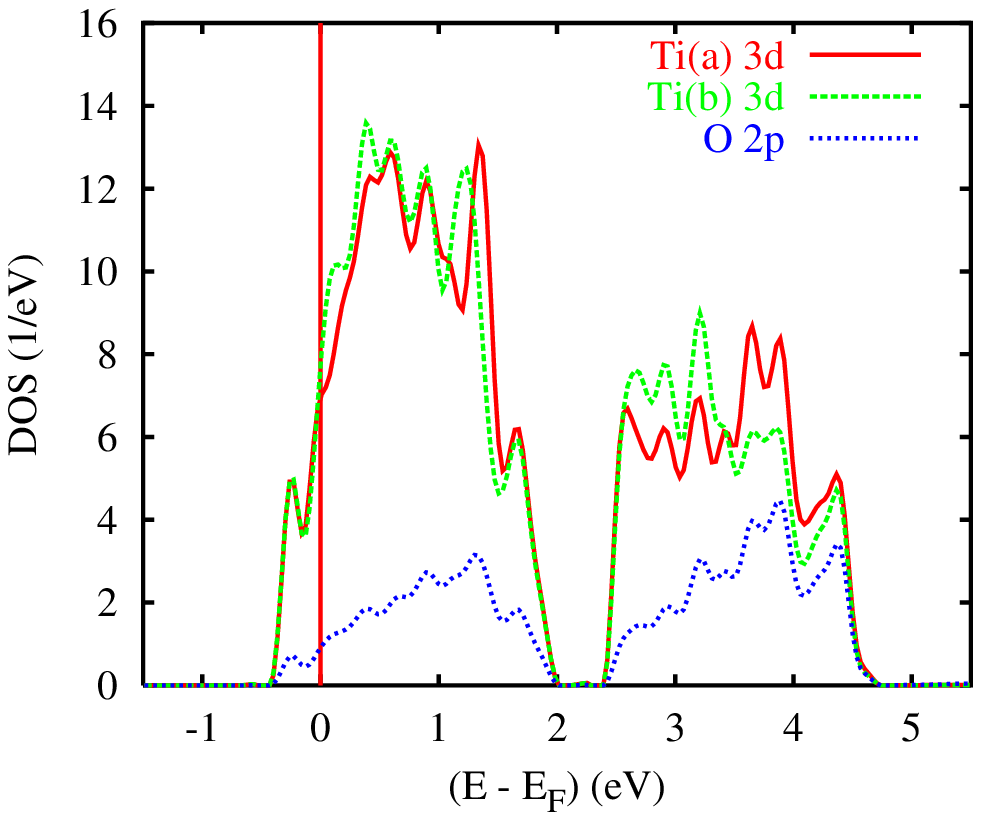}}
\subfigure{\includegraphics[width=0.48\textwidth]{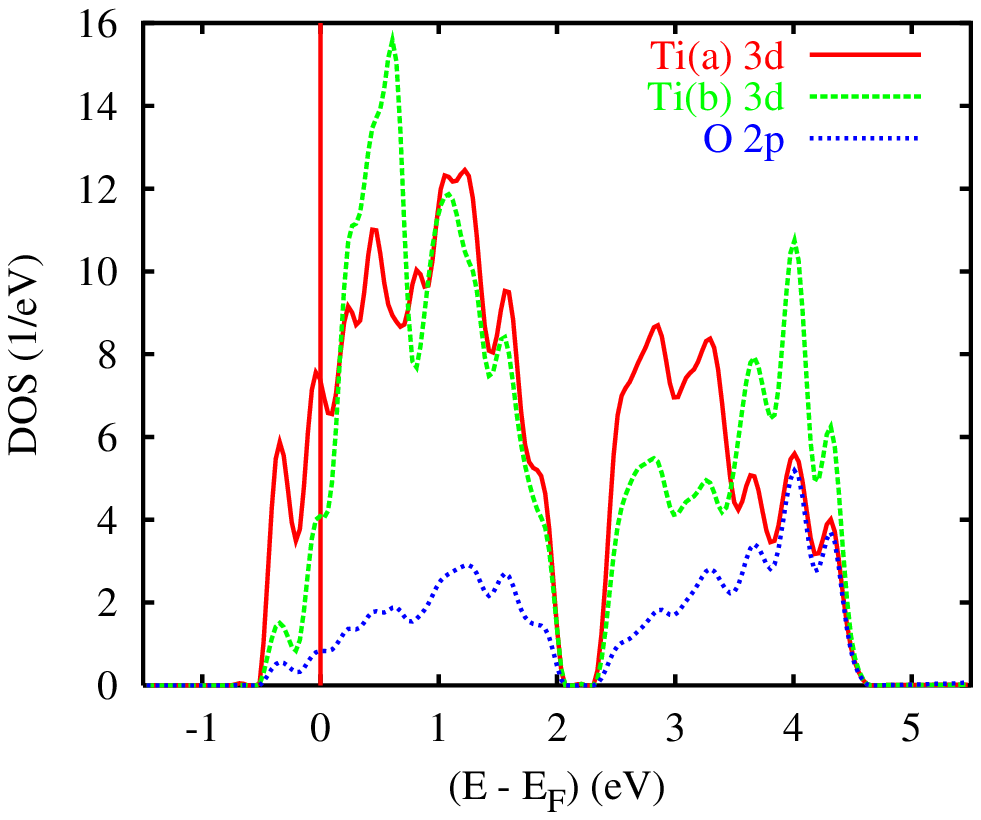}}
\caption{Partial DOS as resulting from the room-temperature (left) 
         and low-temperature (right) crystal structure.
         Here and in the following figures slight broadening is due 
         to the DOS calculation scheme \protect \cite{methfessel89}.}
\label{fig:res1}
\end{figure}

Calculated partial densities of states (DOS) are displayed in Fig.\ 
\ref{fig:res1}. 
Here and in all following figures results for the room-temperature and 
low-temperature structure are given on the left and right, respectively.  
In Fig.\ \ref{fig:res1}, contributions from both the Ti $ 3d $ and O $ 2p $ 
states are included. In addition, Ti contributions from the (a) and (b) 
chains containing Ti1/Ti3 and Ti2/Ti4, respectively, are distinguished. 
All other orbitals play only a negligible role in the energy interval 
shown. 

Two groups of bands are identified, which extend from $ -0.5 $ to 
$ 2.0 $\,eV and from $ 2.4 $ to $ 4.6 $\,eV and comprise 24 and 16 
bands, respectively. They derive mainly from the Ti $ 3d $ states. 
In addition, 42 bands are found in the energy range from $ -8.5 $ 
to $ -3.8 $\,eV, which trace back mainly to the O $ 2p $ states.  
However, $ p $--$ d $ hybridization causes $ d $ and $ p $ 
contributions, respectively, below and above $ -2 $\,eV 
reaching about 15\% especially in the two upper groups. 
Crystal field splitting expected from the fact that the Ti atoms are 
located at the centers of slightly distorted octahedra is observed in 
the Ti $ 3d $ partial DOS. As a detailed analysis of the wave functions 
reveals, the lower group of Ti $ 3d $ derived bands arises almost 
exclusively from states of $ t_{2g} $ symmetry. In contrast, bands 
between $ 2.4 $ to $ 4.6 $\,eV trace back to the $ e_g $ states. The 
energetical separation of the centers of gravity of these two groups 
of about $ 2.7 $\,eV compares well with the value of $ 2.4 $\,eV taken 
from the XAS measurements by Abbate {\em et al.}\ \cite{abbate95}. At 
the same time, the calculated position and width of the O $ 2p $ 
dominated group of bands is in very good agreement with the photoemision 
data of these authors. 

In general, these findings are very similar for the room-temperature 
and the low-temperature structure. In particular, the O $ 2p $ bands 
keep their energetical position and band width, again in agreement 
with the photoemssion data \cite{abbate95}. Differences show up 
on closer inspection of the Ti $ 3d $ dominated states. Most striking 
are the changes occurring in the occupied part of the Ti $ 3d $ $ t_{2g} $ 
bands. While for the room-temperature structure the Ti $ 3d $ partial 
DOS in this energy interval are almost the same for both chains, we 
observe strong increase and decrease of the Ti $ 3d $ partial DOS 
arising from the atoms of chain (a) and (b), respectively, for the 
low-temperature structure indicative of a considerable charge transfer.  
The integrated difference amounts to $ \approx 0.8 $ electrons 
per formula unit. 
At the same time, states in the Ti $ 3d $ $ e_g $ group of bands, i.e.\ 
between $ 2.4 $ and $ 4.6 $\,eV, experience some rearrangement. In 
particular, on going from the room-temperature results to those obtained 
for the low-temperature structure, we find energetical down- and upshift, 
respectively, of the centers of gravity of these bands for chain (a) 
and (b). 

Both these changes can be attributed to the difference of the average 
Ti-O distances emerging in the low-temperature phase \cite{marezio72}.
They lead to the assignment of Ti $ d^{0} $ and $ d^{1} $ charges 
contrasting the formal $ d^{0.5} $ valence of the room-temperature phase 
\cite{marezio72}. While in the latter phase 
the average Ti-O bond length is the same for all titanium sites, in 
the low-temperature phase the average bond length increases and decreases 
for chain (a) and (b) atoms, respectively. As a consequence, $ d $ electron 
charge is transferred from atoms Ti2/Ti4 to Ti1/Ti3, leaving the $ d $ 
states in the (b) chains unoccupied. In addition, the shrinking and 
inflating of the oxygen octahedra strongly affect the bonding-antibonding 
splitting especially of the $ \sigma $-bonding states of $ e_g $ symmetry. 
This splitting decreases/increases in chains (a)/(b), leading to the 
observed down-/upshift of the Ti $ 3d $ dominated antibonding states. 

\begin{figure}[htb]
\centering 
\subfigure{\includegraphics[width=0.48\textwidth]{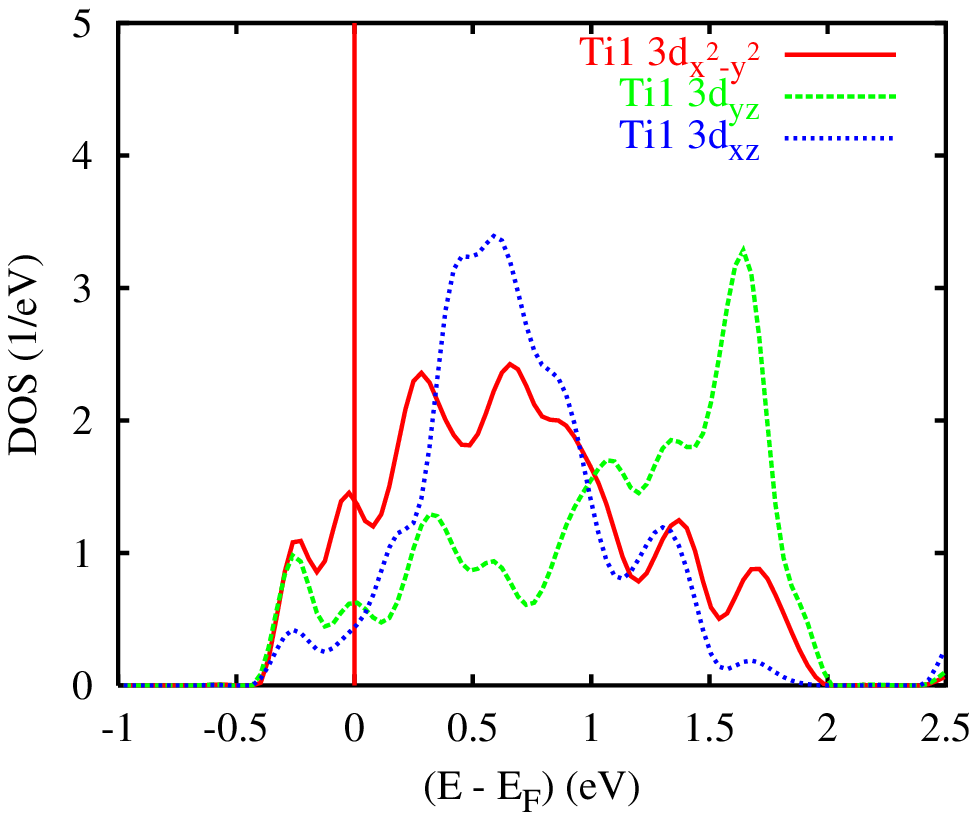}}
\subfigure{\includegraphics[width=0.48\textwidth]{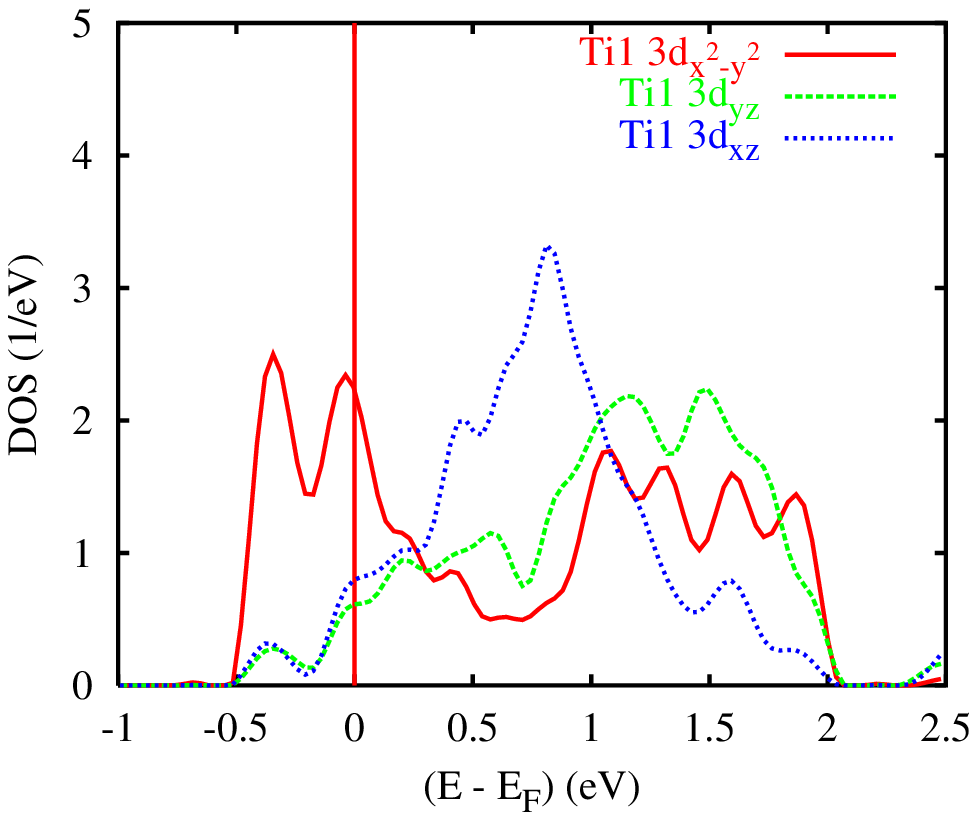}} \\
\subfigure{\includegraphics[width=0.48\textwidth]{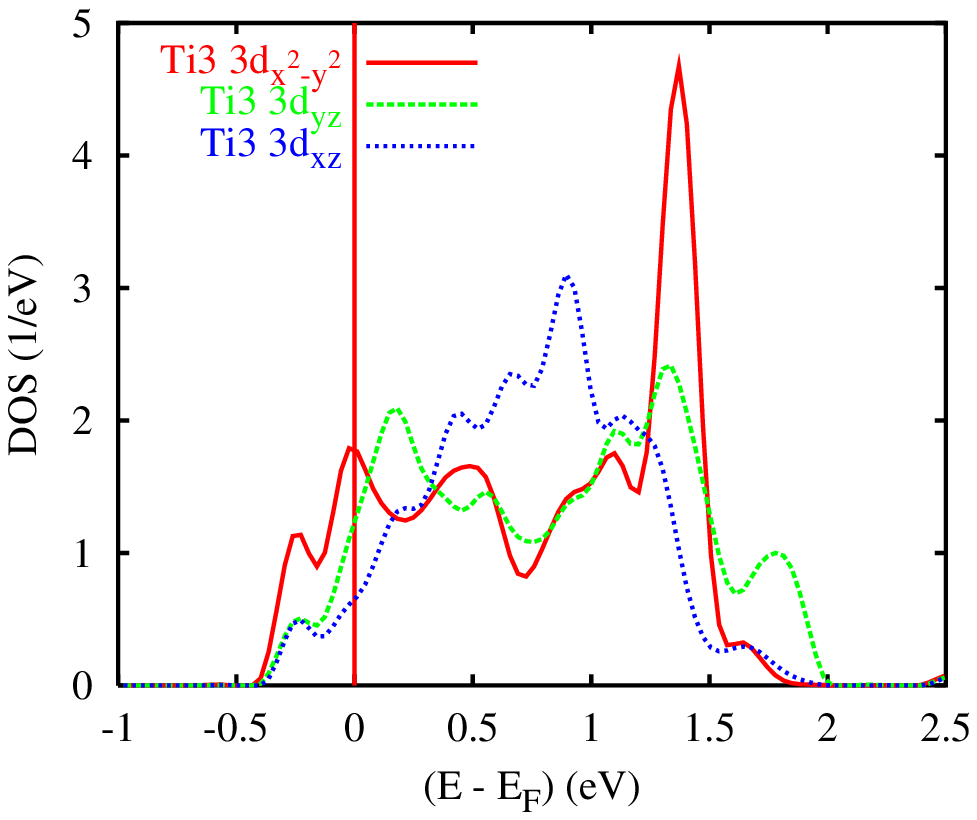}}
\subfigure{\includegraphics[width=0.48\textwidth]{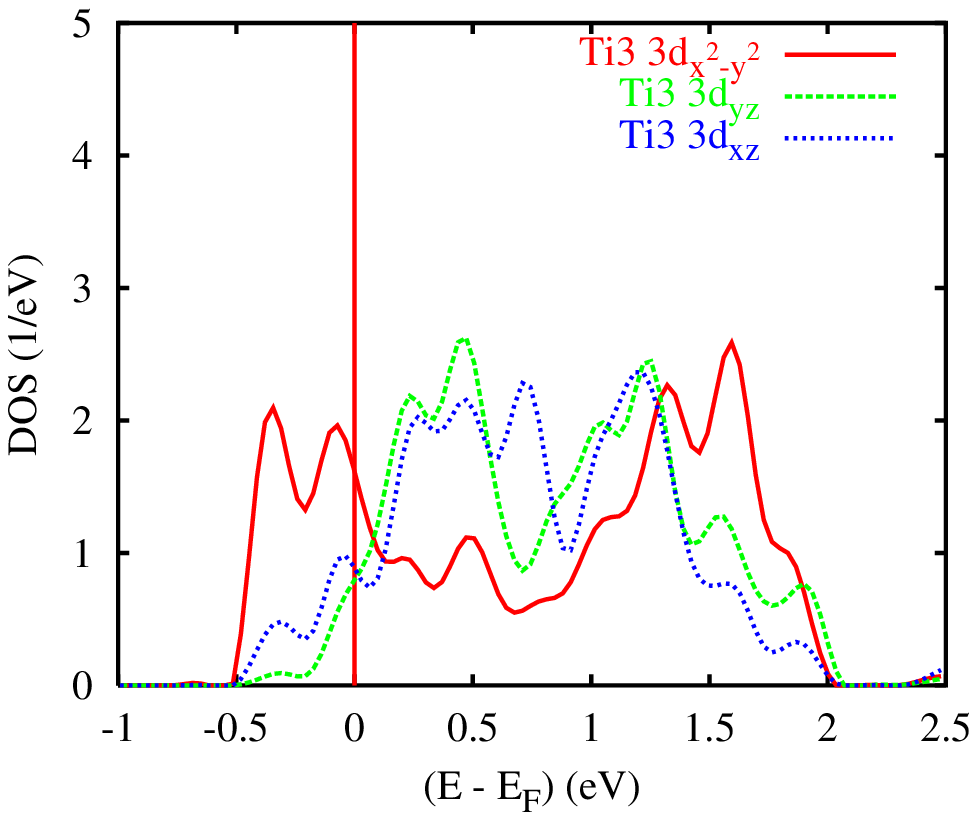}}
\caption{Partial $ 3d $ $ t_{2g} $ DOS of (a) chain atoms Ti1 and Ti3.}
\label{fig:res2}
\end{figure}

\begin{figure}[htb]
\centering 
\subfigure{\includegraphics[width=0.48\textwidth]{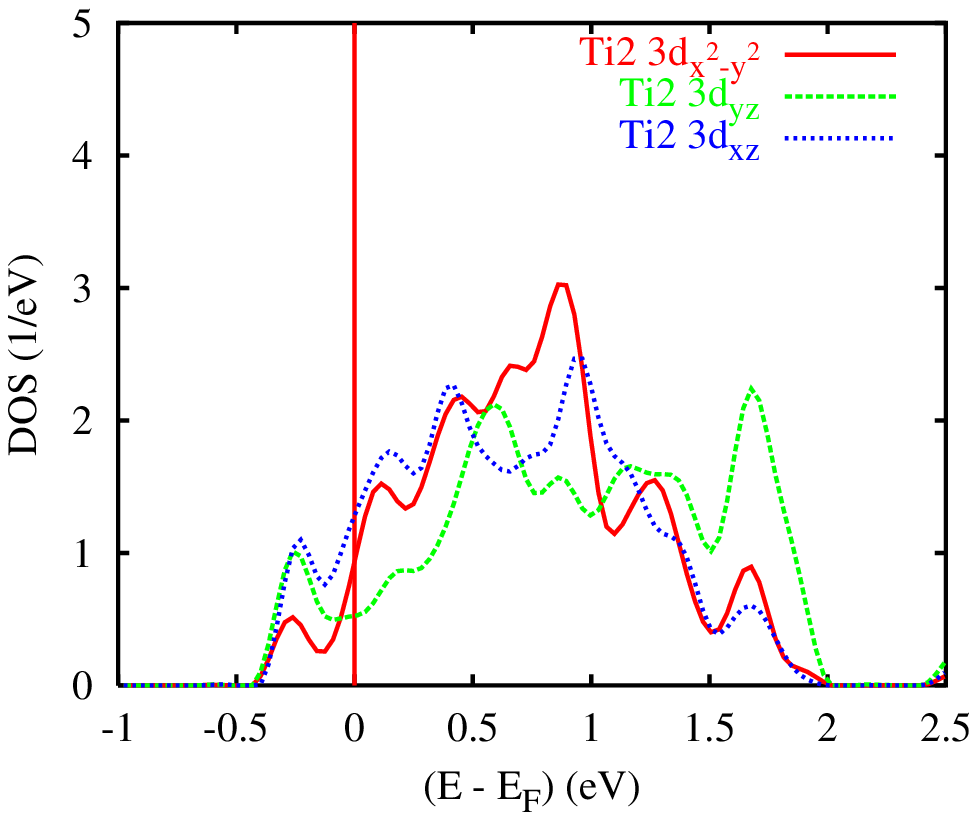}}
\subfigure{\includegraphics[width=0.48\textwidth]{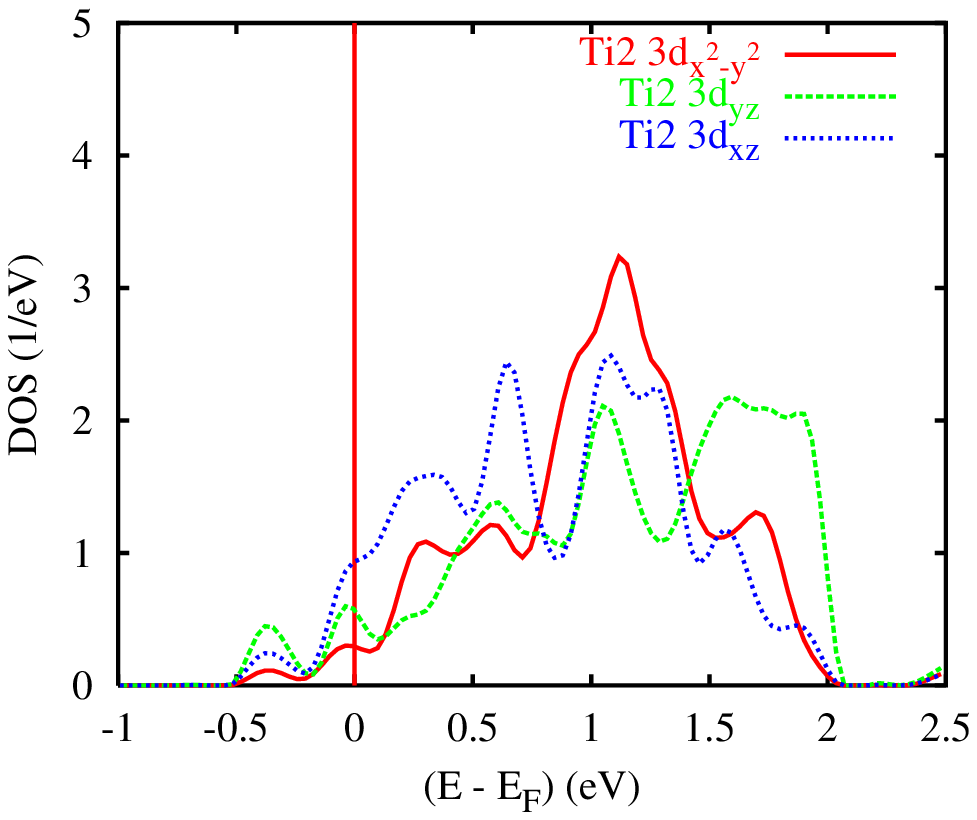}} \\
\subfigure{\includegraphics[width=0.48\textwidth]{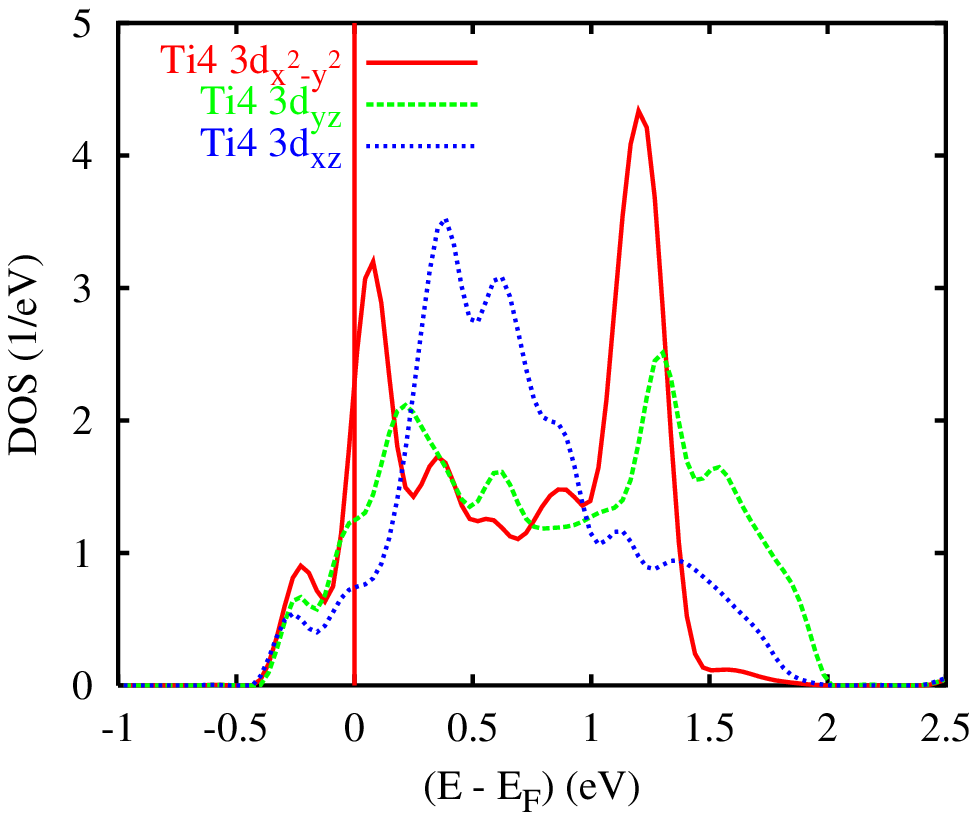}}
\subfigure{\includegraphics[width=0.48\textwidth]{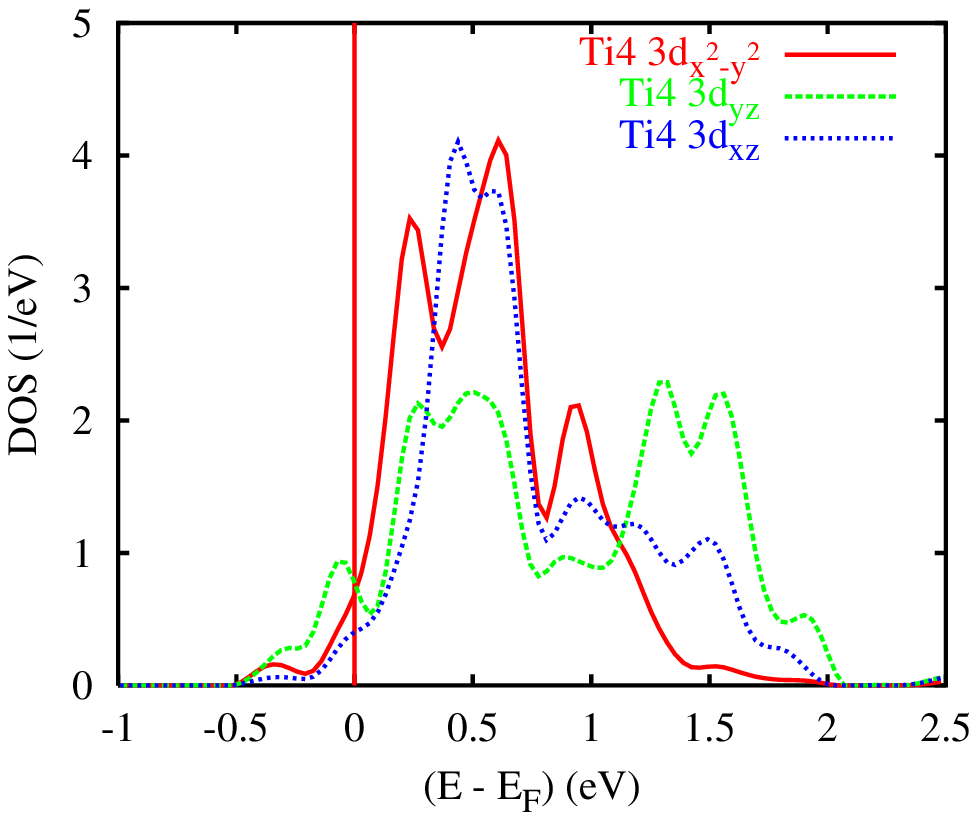}}
\caption{Partial $ 3d $ $ t_{2g} $ DOS of (b) chain atoms Ti2 and Ti4.}
\label{fig:res3}
\end{figure}

In order to investigate the charge redistribution coming with the 
transitions in more detail we display the Ti $ 3d $ $ t_{2g} $ partial 
DOS of all four titanium atoms separately in Figs.\ \ref{fig:res2} 
and \ref{fig:res3}. 
In doing so we distinguish the three different $ d $ states contributing 
to this group of bands using the aforementioned local coordinate system 
as defined in Ref.\ \cite{eyert02b}. In this system, the $ d_{x^2-y^2} $ 
states, pointing parallel to the pseudorutile $ c_{\rm prut} $ axis, 
mediate $ \sigma $-type metal-metal overlap within the chains. In contrast, 
the $ d_{yz} $ orbitals take part in albeit weaker $ \sigma $-type 
$ d $-$ d $ overlap across the chains. Weak interchain coupling is also 
due to overlap of $ d_{x^2-y^2} $ and $ d_{xz} $ orbitals of neighbouring 
chains. 

For the high-temperature structure strong splitting of the $ d_{x^2-y^2} $ 
states of both Ti3 and Ti4 is observed. This is a consequence of the 
short Ti3-Ti3 and Ti4-Ti4 distances in the chain centers causing strong 
bonding-antibonding splitting of the $ d_{x^2-y^2} $ orbitals of these 
atoms. In contrast, distances between the chain center and end atoms are 
considerably larger, this leading to the somewhat unstructured shape of 
the $ d_{x^2-y^2} $ partial DOS of atoms Ti1 and Ti2. Similar shapes are  
displayed by the $ d_{xy} $ partial DOS of all atoms due to the very small  
metal-metal bonding these states take part in. 
Finally, the $ d_{yz} $ states of all atoms 
show distinct splitting into bonding and antibonding states, reflecting 
overlap of these orbitals of atoms Ti3 and Ti4 (peaks at $ 0.2 $ and 
$ 1.35 $\,eV) as well as atoms Ti1 and Ti2. In the latter case overlap 
is parallel to the pseudorutile $ a_{\rm prut} $ axis perpendicular to the 
chains across octahedral faces. This bond corresponds to the $ c $-axis 
pairs in the sesquioxide. Indeed, the shape of the $ d_{yz} $ partial DOS 
with its pronounced antibonding peak at $ 1.6 $\,eV and the tail of bonding 
states extending from $ -0.3 $ to $ 1.3 $\,eV has been also observed for 
$ {\rm V_2O_3} $. 

On going to the low-temperature phase several striking changes occur. 
First, we observe the abovementioned increase and decrease of the partial 
DOS of the (a) and (b) chain atoms, respectively, in the occupied part 
of the spectrum, which is a consequence of the charge ordering coming 
with the increased and reduced average Ti-O bond-length as reported by 
Marezio and coworkers \cite{marezio72,marezio73}. In the (b) chains 
smaller distances induce a stronger bonding-antibonding splitting of 
the hybridized Ti $ 3d $ and O $ 2p $ states, hence, upshift of the 
antibonding $ d $ dominated bands. This effect is supported by the 
displacement of these atoms perpendicular to the chains. Eventually, 
we witness almost complete depopulation of the (b) chain states, 
i.e.\ Ti $ d^0 $ valence. In contrast, on the (a) chains longer Ti-O 
distances lead to reduced bonding-antibonding splitting, hence, 
energetical lowering of the $ d $ bands and the occupation reaches 
nearly $ d^1 $. 

Second, in addition to the charge ordering we find strong changes in 
the occupation of the single $ d $ states of the (a) chain atoms. Due 
to the albeit small displacement of atoms Ti1 and Ti3 perpendicular 
to the chain axis the $ d_{xz} $ and $ d_{yz} $ states of these atoms 
are subject to larger $ d $-$ p $ hybridization and thus shift to higher 
energies. In contrast, both the Ti1 and Ti3 $ d_{x^2-y^2} $ partial DOS 
show strong splitting into a double maxima structure with peaks near 
$ -0.25 $ and $ 1.5 $\,eV. This effect results from the dimerization of 
these two atoms observed below the metal-insulator transition, which leads 
to metal-metal overlap of the orbitals extending along the (a) chains 
and splitting of the corresponding bands into bonding and antibonding 
states. At the same, chain (b) atoms do not display any splitting of 
these states due to the uniform distance of more than $ 3.0 $\,\AA \ 
between all atoms of this chain. 
 
\begin{figure}[htb]
\centering 
\subfigure{\includegraphics[width=0.48\textwidth]{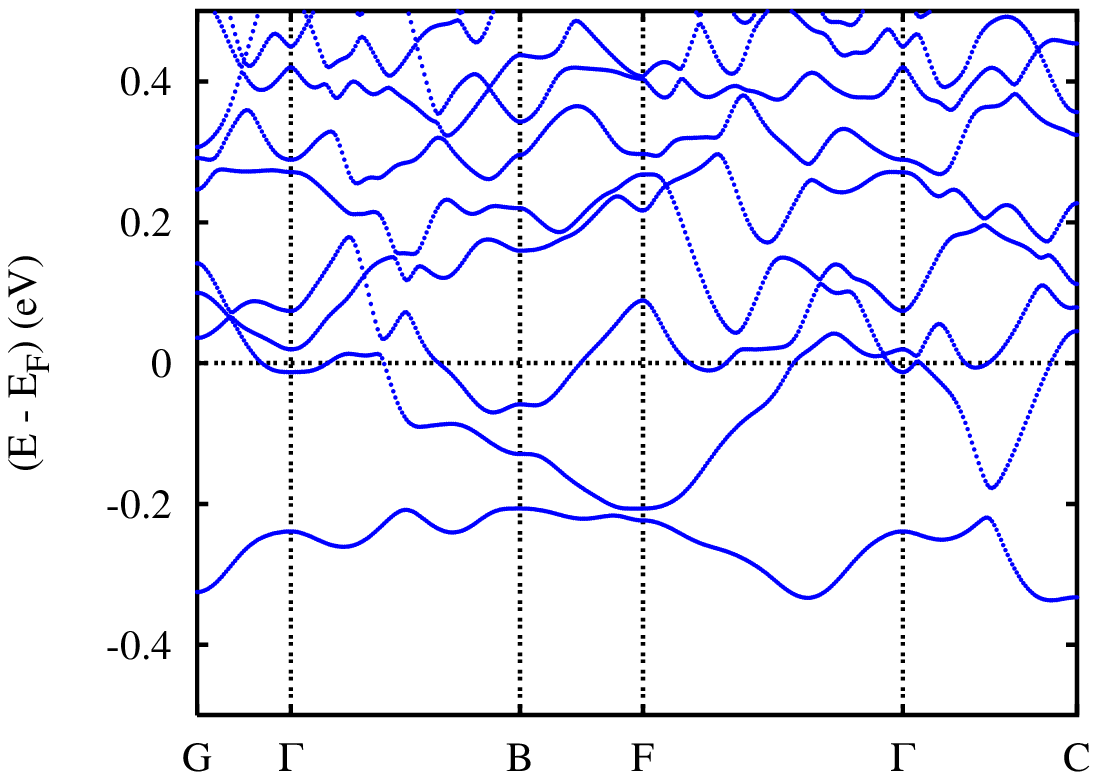}}
\subfigure{\includegraphics[width=0.48\textwidth]{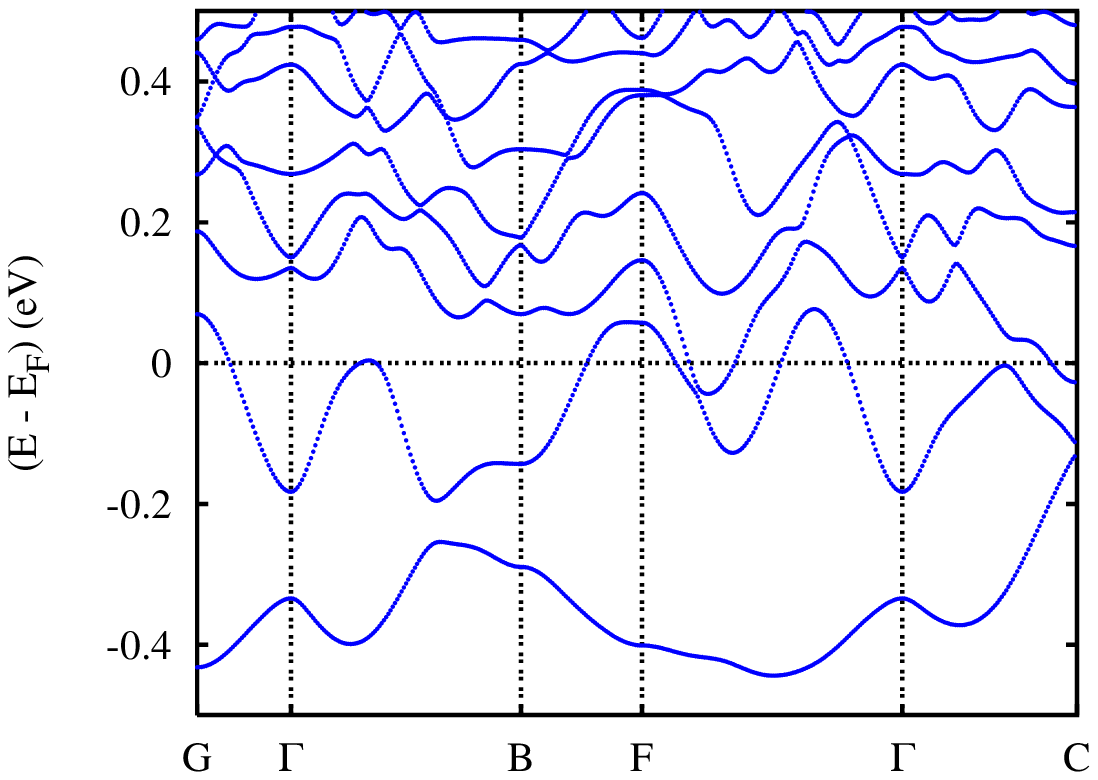}}
\caption{Electronic band structure.} 
\label{fig:res4}
\end{figure}

The corresponding changes of the band structure as shown in Fig.\ 
\ref{fig:res4}
have drastic effects 
especially near the Fermi energy. While in the room-temperature 
structure bands in the lower part of the $ t_{2g} $ group show strong 
dispersion leading to the metallic behaviour, the low-temperature 
structure is characterized by two split-off bands. As expected from 
the partial DOS and as revealed by a detailed analysis of the wave 
functions, these weakly dispersing bands are almost exclusively of 
Ti1 and Ti3 $ d_{x^2-y^2} $ character. Additional bands of the same 
symmetry, which likewise show a reduced dispersion, are found at the 
upper edge of the $ t_{2g} $ group, supporting the previous 
interpretation in terms of bonding-antibonding splitting of these 
states. Yet, despite the separation of the lower $ d_{x^2-y^2} $ bands 
from all higher lying states there is still a semimetal-like overlap 
of about $ 0.1 $\,eV, inhibiting the complete opening of the insulating 
band gap. 

In general, the situation resembles very much that found for the rutile-type 
compounds $ {\rm TiO_2} $ and $ {\rm VO_2} $, where the metal atoms 
are found in a $ d^0 $ and $ d^1 $ configuration, respectively. While 
the insulating titanate preserves the rutile structure down to lowest 
temperatures maintaining especially the equidistant spacing of the 
metal atoms, vanadium dioxide shows distinct deviations from this 
structure at the metal-insulator transition. In particular, vanadium 
atoms experience pairing parallel to the $ c_{\rm rut} $ axis as well as 
zigzag-like antiferroelectric displacements perpendicular to this axis. 
In the calculation for $ {\rm VO_2} $, these changes lead to splitting 
of the $ d_{x^2-y^2} $ bands as well as energetical upshift of the 
$ d_{xz} $ and $ d_{yz} $ states \cite{eyert02b}. We may thus regard 
low-temperature $ {\rm Ti_4O_7} $ as composed of $ {\rm TiO_2} $- and 
$ {\rm VO_2} $-like chains, which display the same behaviour as the 
respective dioxide. Common to $ {\rm VO_2} $ and $ {\rm Ti_4O_7} $ 
is also the finding of a small semimetal-like overlap for the low-temperature 
phase, which we attribute to the shortcomings of the local density 
approximation. Yet, this weakness does not undermine the interpretation 
of the phase transitions of both compounds as arising from orbital 
order and strong bonding-antibonding splitting of the $ d_{x^2-y^2} $ 
bands due to the structural changes and strong electron-phonon 
coupling. In addition, $ {\rm Ti_4O_7} $ is characterized by distinct 
charge ordering, leading to separated $ d^0 $ and $ d^1 $ chains and 
laying ground for the efficacy of the aforementioned mechanisms.

\section{Conclusion}

In summary, according to electronic structure calculations for the 
Magn\'{e}li phase $ {\rm Ti_4O_7} $, the metal-insulator transition 
of this material arises from the complex interplay of different types of 
ordering phenomena. While in the room-temperature structure all Ti $ 3d $ 
$ t_{2g} $ states display uniform occupation, the low-temperature structure 
is characterized by the formation of two types of titanium chains comprising 
Ti $ d^0 $ and $ d^1 $ states, respectively. This charge ordering is 
accompanied by orbital ordering on the $ d^1 $ chains and strong 
bonding-antibonding splitting of the respective $ d_{x^2-y^2} $ states 
coming with Ti-Ti dimerization and zigzag-type antiferroelectric 
displacement of these atoms in complete accordance with the findings 
for the $ d^1 $ dioxide $ {\rm VO_2} $. Finally, like in the latter 
compound, singlet formation on the $ d_{x^2-y^2} $ states paves the way 
for the observed insulating behaviour. 

To conclude, $ {\rm Ti_4O_7} $ shares fundamental aspects with other 
cornerstone materials of condensed matter physics. The charge ordering 
occuring at the metal-insulator transition is closely related to that 
discussed for the Verwey transition of $ {\rm Fe_3O_4} $. In contrast, 
the orbital order and metal-metal dimerization accompanying the phase 
transition of $ {\rm Ti_4O_7} $ arise from the same mechanism as that 
driving the transitions of $ {\rm VO_2} $ and $ {\rm NbO_2} $, namely 
instability of the $ d^1 $ atom chains towards the strong metal-metal 
bonding along the characteristic chains as well as antiferroelectric 
displacement perpendicular. This effect may thus be regarded as a 
universal mechanism common to many early rutile-type transition-metal 
compounds.

\section*{Acknowledgments}
This work was supported by the Deutsche Forschungsgemeinschaft (DFG) through 
Sonderforschungsbereich SFB 484.

\end{document}